# Field-Free Reconfigurable Spin Logic in Compositionally Graded $Mn_xCoAl$ Layer

*Haohang Cheng[1,2,5], Pengju Wang[3], Hongrui Qin[1], Rongkun Han[1,4*], Dong Pan[1,2], Dahai Wei[1,2], and Jianhua Zhao[1,3*]*

1 State Key Laboratory of Semiconductor Physics and Chip Technologies, Institute of Semiconductors, Chinese Academy of Sciences, Beijing 100083, China

2 School of Materials Science and Optoelectronic Technology, University of Chinese Academy of Sciences, Beijing 100049, China

3 State Key Laboratory of Spintronics, Hangzhou International Innovation Institute, Beihang University, Hangzhou 311115, China

4 Institut Jean Lamour, Université de Lorraine, CNRS, UMR 7198, Nancy, France

5 Laboratory of Solid-State Optoelectronics Information Technology, Institute of Semiconductors, Chinese Academy of Sciences, Beijing 100083, China

*Corresponding authors: Jianhua Zhao (jianhuazhao@buaa.edu.cn); Rongkun Han (rongkun.han@univ-lorraine.fr)

---

Spintronic logic-in-memory provides a route to reducing data-transfer energy consumption and accelerating computation by integrating nonvolatile storage and logic functions within a single physical unit. Here, we investigate field-free spin-orbit-torque (SOT) switching and reconfigurable spin logic in a vertically composition-graded $Mn_xCoAl$/Pt heterostructure. In contrast to the uniform-composition $Mn_{1.8}CoAl$/Pt and $Mn_{2.5}CoAl$/Pt control devices, the $Mn_{1.8\text{-}2.5}CoAl$ gradient device exhibits deterministic SOT switching under zero external magnetic field. Dzyaloshinskii-Moriya interaction (DMI)-related measurements reveal a finite characteristic effective-field scale, indicating the involvement of chiral magnetization-reversal processes. Meanwhile, the vertical composition gradient introduces magnetic inhomogeneity that can modify domain nucleation and propagation, and the combined effects are consistent with the observed field-free switching behavior. The composition-gradient device further exhibits accumulative multistate switching with stable intermediate Hall-resistance states under zero magnetic field, whereas the uniform-composition controls do not show comparable stable multistate behavior. By tailoring the amplitude, polarity, and sequence of current pulses, the device can be tuned between binary and multilevel switching modes, enabling the experimental realization of AND, OR, NAND, and NOR Boolean operations within the same Hall-bar device without any external magnetic field. These results extend Mn-Co-Al Heusler heterostructures from field-assisted SOT switching toward field-free magnetization control, multistate operation, and reconfigurable spin logic through vertical composition engineering.

# 1 INTRODUCTION

The rapid growth of data-intensive computing has increasingly exposed the limitations of conventional von Neumann architectures, in which the physical separation between memory and processing units results in substantial data-transfer overhead and latency.[1–3] Logic-in-memory (LIM), which combines information storage and computation within the same physical hardware, provides an alternative route toward more compact and nonvolatile computing architectures.[1,3] Magnetic random-access memory (MRAM) and spin-based logic are attractive candidates because of their nonvolatility, high endurance, and compatibility with electrical write and read operations.[4,5] Among the available electrical writing mechanisms, spin-orbit torque (SOT) enables current-driven manipulation of magnetization through charge-to-spin conversion associated with the spin Hall effect and interfacial spin-orbit coupling.[6–9] Magnetic materials with robust perpendicular magnetic anisotropy (PMA), high spin polarization, and low Gilbert damping are ideal candidates for fabricating SOT devices.[5] Mn-Co-Al-based Heusler alloys are promising in this respect because of their highly spin-polarized electronic structures [10,11] and composition-dependent magnetic anisotropy.[12] In particular, SOT-induced magnetization switching has previously been demonstrated in Pt/$Mn_{1+x}Co_{2-x}Al$ heterostructures under assisting in-plane magnetic fields, establishing the feasibility of electrical manipulation in this Heusler material family.[12]

For conventional heavy-metal/ferromagnet bilayers, deterministic field-free SOT switching generally requires an additional symmetry-breaking mechanism. Representative approaches include exchange bias from antiferromagnetic layers,[13] laterally asymmetric structures,[14] wedge-shaped structures,[15] magnetic multilayers,[16–19] interlayer Dzyaloshinskii–Moriya interaction,[20] and unconventional spin currents with out-of-plane spin polarization.[21,22] These studies have demonstrated several approaches to achieving deterministic perpendicular SOT switching without an external magnetic field.[23–26] Vertical composition grading offers another route to field-free SOT switching, as demonstrated in CoTb as well as in the Heusler alloy MnPtGe.[27–29] In Mn-Co-Al Heusler alloys, previous studies have mainly focused on magnetic properties and field-assisted SOT switching,[12,30] while field-free switching and composition-gradient engineering remain less explored. Moreover, SOT has also been widely explored for programmable and reconfigurable spin logic. Various SOT-based devices have demonstrated Boolean logic, domain-wall logic, multistate memory, and reconfigurable logic functions,[31–35] highlighting the potential of SOT for integrated nonvolatile logic-in-memory applications.

In this work, we extend the SOT functionality of $Mn_xCoAl$/Pt Heusler heterostructures to field-free operation, multistate control, and reconfigurable spin logic. We compare a vertically graded $Mn_{1.8\text{-}2.5}CoAl$ structure with uniform $Mn_{1.8}CoAl$ and $Mn_{2.5}CoAl$ control samples. The composition-gradient device exhibits field-free SOT switching, whereas the uniform-composition controls show no deterministic zero-field reversal. Dzyaloshinskii-Moriya interaction (DMI)-related measurements reveal a finite characteristic effective-field scale ($H_{DMI}$) in the gradient device, indicating the involvement of chiral magnetization-reversal processes in the field-free switching. Furthermore, the composition-gradient device exhibits accumulative multistate switching with stable intermediate Hall-resistance states, whereas the uniform-composition controls do not show comparable stable multistate behavior. Based on this field-free multistate response, AND, OR, NAND, and NOR Boolean operations are realized within the same Hall-bar device under zero external magnetic field. These results demonstrate the integration of field-free SOT switching, multistate operation, and reconfigurable spin logic in composition-engineered Mn-Co-Al Heusler heterostructures.

# 2 EXPERIMENTAL SECTION

The epitaxial $Mn_xCoAl$ (5 nm)/Pt (5 nm) layers were grown on GaAs (100) substrates by molecular-beam epitaxy [Fig. 1(a)]. The 5-nm-thick $Mn_xCoAl$ layer was deposited at a substrate temperature of 250 °C. The Co and Al fluxes were maintained constant throughout the growth, whereas the Mn flux was continuously varied by controlling the Mn effusion-cell temperature. A 5-nm-thick Pt capping layer was subsequently deposited at room temperature to prevent oxidation and serve as the spin-current source. Three samples were prepared: uniform $Mn_{1.8}CoAl$ and $Mn_{2.5}CoAl$, and a composition-gradient $Mn_{1.8\text{-}2.5}CoAl$ sample, where the subscripts of the gradient sample indicate the nominal Mn composition from the GaAs interface to the Pt interface. The thin films were patterned into Hall bar devices with a channel width of 10 μm using standard ultraviolet photolithography combined with argon ion etching.

The crystal structures of the samples were characterized by XRD using Cu Kα radiation. Magnetic hysteresis loops were measured at room temperature utilizing a vibrating sample magnetometer. Electrical transport measurements and SOT driven magnetization switching experiments were performed using a physical property measurement system (PPMS). For the switching measurements, 50 μs current pulses were applied to drive the magnetization, and the anomalous Hall resistance was recorded after a 5 s delay using a low sensing current of 0.1 mA to ensure data stability and minimize thermal interference.

# 3 RESULTS AND DISCUSSION

### 3.1 Structural Characteristics and Magnetic Properties

Figure 1(b) displays the X-ray diffraction (XRD) patterns of the uniform $Mn_{1.8}CoAl$ and $Mn_{2.5}CoAl$ samples together with the composition-gradient $Mn_{1.8\text{-}2.5}CoAl$ sample. For the $Mn_xCoAl$ magnetic layer, the (004) diffraction peak exhibits a pronounced composition dependence, shifting from 61.98° for $Mn_{1.8}CoAl$ to 59.06° for $Mn_{2.5}CoAl$. The diffraction peak of the $Mn_{1.8\text{-}2.5}CoAl$ gradient sample is located between those of the two uniform-composition samples, indicating that its average out-of-plane lattice parameter lies within the range defined by $Mn_{1.8}CoAl$ and $Mn_{2.5}CoAl$. Correspondingly, the lattice constant expands significantly from 5.99 to 6.22 Å as the Mn content increases from 1.8 to 2.5. Figure 1(c) presents the out-of-plane magnetic hysteresis loops of the three $Mn_xCoAl$ films measured at 300 K. All samples maintain robust perpendicular magnetic anisotropy (PMA) despite the substantial variation in Mn composition. The saturation magnetization $M_s$ decreases from $Mn_{1.8}CoAl$ to $Mn_{2.5}CoAl$, which is attributed to the enhanced antiferromagnetic coupling associated with excess Mn atoms.[12] The magnetic response of the $Mn_{1.8\text{-}2.5}CoAl$ gradient sample lies between those of the two uniform-composition samples. The anomalous Hall effect (AHE) curves shown in Figure 1(d) further confirm the presence of PMA in all three samples, with the coercive field of the gradient sample also lying between those of $Mn_{1.8}CoAl$ and $Mn_{2.5}CoAl$. The PMA in $Mn_xCoAl$ originates from tetragonal lattice distortion,[12] where the incorporation of excess Mn modifies the out-of-plane lattice constant and the $c/a$ ratio, thereby affecting the spin-orbit-coupling-induced magnetic anisotropy. Consequently, the composition-dependent evolution of the lattice distortion, together with the reduction in $M_s$, results in the pronounced variation in magnetic properties with Mn composition.

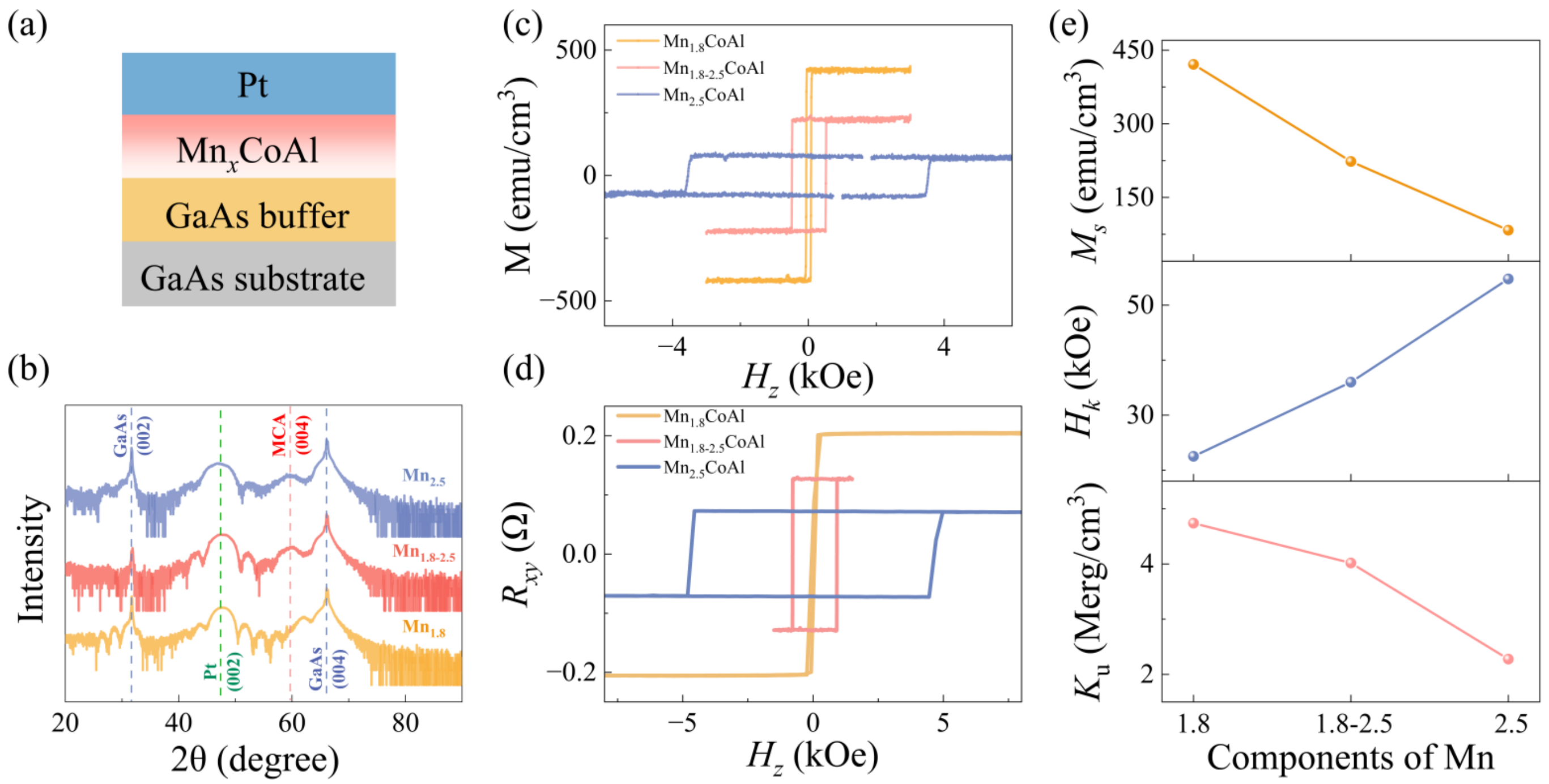

**Figure 1.** (a) Schematic illustration of the GaAs/$Mn_xCoAl$/Pt heterostructure with a vertical Mn-composition gradient. (b) XRD θ-2θ scans of the uniform $Mn_{1.8}CoAl$ and $Mn_{2.5}CoAl$ samples and the composition-gradient $Mn_{1.8-2.5}CoAl$ sample. (c) Out-of-plane magnetic hysteresis (*M-H*) loops. (d) AHE resistance ($R_{xy}$) curves measured under a perpendicular magnetic field $H_z$. (e) Comparison of $M_s$, $H_k$, and $K_u$ on the Mn composition $x$.

Figure 1(e) summarizes the saturation magnetization $M_s$, effective anisotropy field $H_k$, and magnetic anisotropy constant $K_u$ of the three $Mn_xCoAl$ samples. For the two uniform-composition samples, $M_s$ decreases while $H_k$ increases as the Mn content increases from $x$=1.8 to $x$=2.5. The $Mn_{1.8-2.5}CoAl$ gradient sample exhibits intermediate $M_s$ and $H_k$ values between those of $Mn_{1.8}CoAl$ and $Mn_{2.5}CoAl$, consistent with its intermediate magnetic characteristics observed in the hysteresis measurements. Detailed measurement and calculation procedures are provided in Supplementary Note 1. The magnetic anisotropy constant $K_u$ is calculated by:

$$K_u=\frac{1}{2}M_sH_k \tag{1}$$

For the uniform-composition samples, although $H_k$ increases with increasing Mn content, the pronounced reduction in $M_s$ leads to an overall decrease in $K_u$. The $K_u$ value of the $Mn_{1.8-2.5}CoAl$ gradient sample also lies between those of $Mn_{1.8}CoAl$ and $Mn_{2.5}CoAl$. These results demonstrate that the magnetic properties of $Mn_xCoAl$ can be tuned through composition engineering, while the vertically graded $Mn_{1.8-2.5}CoAl$ structure retains robust PMA and overall magnetic characteristics within the range defined by the two uniform-composition reference samples.

### 3.2 Field-Free SOT-Driven Switching

The current-driven magnetization switching is investigated in the three samples. Figure 2(b) compares the current-induced switching behaviors of the three devices under zero external magnetic field. The composition-gradient $Mn_{1.8\text{-}2.5}CoAl$ device shows field-free SOT switching, whereas neither of the two uniform-composition devices exhibits deterministic switching. The $Mn_{1.8\text{-}2.5}CoAl$ device exhibits a pronounced switching response with a maximum switching ratio of 48.9%. For quantitative evaluation, the critical switching current $I_c$ is defined as the applied pulse current at which 50% of the full anomalous-Hall switching amplitude is reached. The current flowing through the Pt layer is estimated by accounting for the current partition between the two layers. The critical current density is then calculated using the Pt current and the Pt cross-sectional area, yielding $4.9 \times 10^7$ A/cm$^2$ for the composition-gradient device. Details are provided in Supplementary Note 2. These results indicate that the vertical composition gradient plays an important role in regulating the field-free current-induced magnetization switching.

The effective symmetry-breaking field is evaluated by measuring the SOT switching curves under different in-plane magnetic fields $H_x$, as shown in Figure 2(c). The switching polarity changes as $H_x$ crosses a small negative compensation field, while the switching amplitude is strongly suppressed near this point. For the $Mn_{1.8\text{-}2.5}CoAl$ device, the polarity reversal occurs near $H_x$≈-23 Oe, indicating the presence of an effective symmetry-breaking bias involved in the zero-field switching.

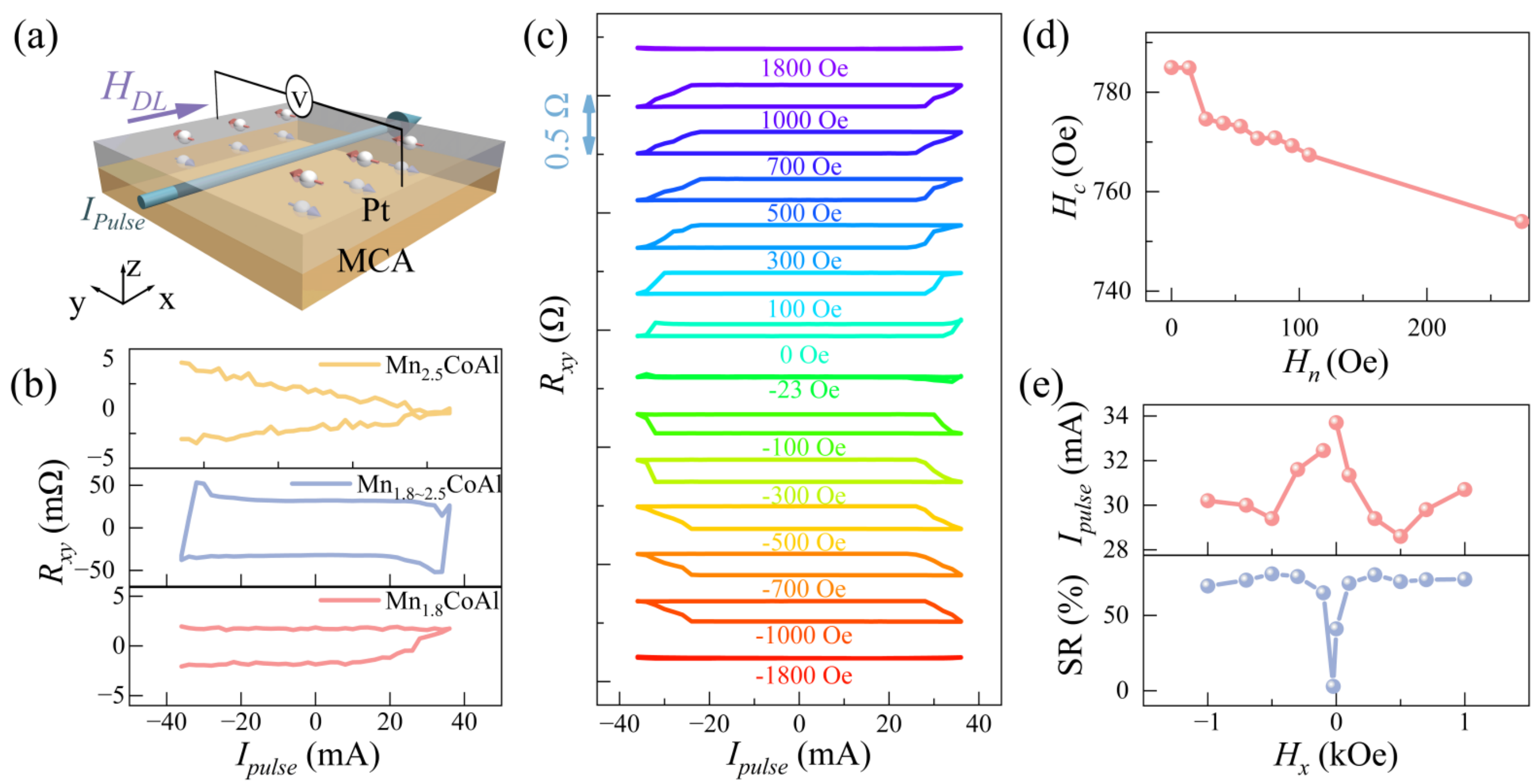

**Figure 2.** (a) Schematic illustration of the Hall-bar device and the measurement configuration. (b) $R_{xy}$ as a function of the pulse current $I_{pulse}$ measured under zero external magnetic field for the uniform $Mn_{1.8}CoAl$ and $Mn_{2.5}CoAl$ samples and the composition-gradient $Mn_{1.8\text{-}2.5}CoAl$ sample. (c) SOT switching curves of the $Mn_{1.8\text{-}2.5}CoAl$ device measured under various in-plane magnetic fields $H_x$. (d) $H_c$ as a function of $H_n$ for the $Mn_{1.8\text{-}2.5}CoAl$ sample, used to evaluate $H_{DMI}$. (e) Dependence of the critical switching current $I_{pulse}$ and switching ratio (SR) on $H_x$.

To further examine the magnetic interaction involved in the reversal process, the $H_{DMI}$ of the $Mn_{1.8-2.5}CoAl$ sample is evaluated in Figure 2(d). Here, $H_c$ denotes the out-of-plane coercive-field component, while $H_n$ denotes the accompanying in-plane field component obtained from the angular-field measurements. $H_{DMI}$ is evaluated from the dependence of $H_c$ on $H_n$ using the magnetic droplet nucleation model,[36] as described in Supplementary Note 3. The characteristic $H_{DMI}$ is estimated to lie in the range of approximately 13-27 Oe. The finite $H_{DMI}$ indicates that DMI-related chiral magnetization-reversal processes are involved in the switching behavior.

DMI introduces chirality into the magnetization-reversal process,[36] while the vertical composition gradient introduces magnetic inhomogeneity that can modify domain nucleation and propagation. Together with the absence of deterministic zero-field switching in the uniform-composition controls, these observations are consistent with a cooperative role of DMI-related chiral reversal and composition-gradient-induced magnetic inhomogeneity in facilitating field-free switching.

Figure 2(e) summarizes the dependence of the switching ratio and critical switching current on the applied in-plane magnetic field $H_x$. The switching ratio is defined as SR=$|\Delta R_{xy}|/R_{AHE}$, where $\Delta R_{xy}$ represents the Hall-resistance change induced by current-driven magnetization reversal. As $H_x$ increases, the switching ratio exhibits a nonmonotonic dependence, initially increasing and subsequently decreasing. At moderate $H_x$, the in-plane field assists the magnetization-reversal process and enhances the accessible switching amplitude. With further increasing $H_x$, however, the magnetization is progressively tilted toward the film plane, weakening the perpendicular magnetic state and suppressing the current-induced out-of-plane reversal. The critical switching current exhibits a correlated nonmonotonic dependence on $H_x$, initially decreasing as the in-plane field lowers the reversal barrier and increasing again at larger $H_x$ as the magnetization is progressively tilted toward the film plane.

### 3.3 Controllable Multistate Nonvolatile Memory

We further investigate the potential of the composition-gradient device for analog neuromorphic computing. Under zero external magnetic field, the $Mn_{1.8-2.5}CoAl$ device exhibits accumulative multistate switching with stable intermediate Hall-resistance states. Under consecutive current pulses, the device exhibits clear accumulative multistate switching with stable intermediate resistance states, as shown in Figure 3(a).

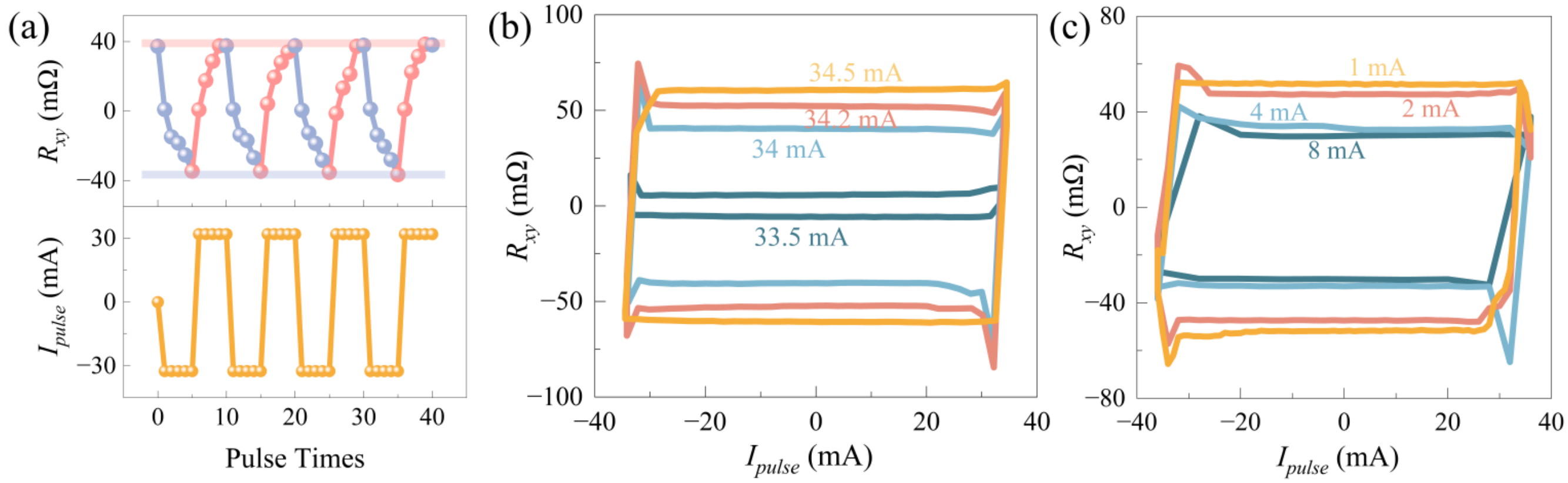

**Figure 3.** (a) Multistate SOT switching behavior driven by consecutive current pulses of +32 mA and -32.5 mA. (b) Dependence of $R_{xy}$ on the applied pulse current $I_{pulse}$ measured under various maximum current amplitudes. (c) $R_{xy}$ as a function of the pulse current $I_{pulse}$ measured with different current step sizes up to a constant maximum amplitude of 36 mA. All measurements are performed under zero external magnetic field.

For comparison, field-assisted control measurements were performed on the uniform-composition devices under an in-plane field of -300 Oe. The uniform $Mn_{1.8}CoAl$ device exhibits abrupt binary switching without stable intermediate resistance states, whereas the uniform $Mn_{2.5}CoAl$ device does not show deterministic switching, as shown in Supplementary Note 4. Together with the multistate behavior observed in the composition-gradient device, these control measurements indicate that the vertical composition gradient plays an important role in promoting accumulative multistate magnetization reversal. By optimizing the pulse conditions to +32 mA and -32.5 mA, the $Mn_{1.8\text{-}2.5}CoAl$ device exhibits multiple stable intermediate states, with a field-free switching amplitude of approximately 30.3%. This nonvolatile resistance modulation resembles synaptic potentiation and depression,[37] indicating its potential for analog neuromorphic and synaptic-device applications.

The effects of pulse parameters are systematically investigated to achieve better control over the multistate switching. Figure 3(b) shows switching curves obtained with maximum pulse amplitudes increasing from 33.5 mA to 36 mA at a fixed current step of 2 mA. The switching amplitude increases rapidly between 33.5 mA and 34.5 mA and then approaches saturation. This result identifies the 33.5 mA to 34.5 mA range as the optimal range for achieving multistate control. At a fixed maximum current of 36 mA, the field-free switching amplitude decreases as the current step size increases from 1 to 8 mA, as shown in Figure 3(c). By adjusting these pulse parameters, a maximum zero-field switching ratio of 48.9% is achieved.

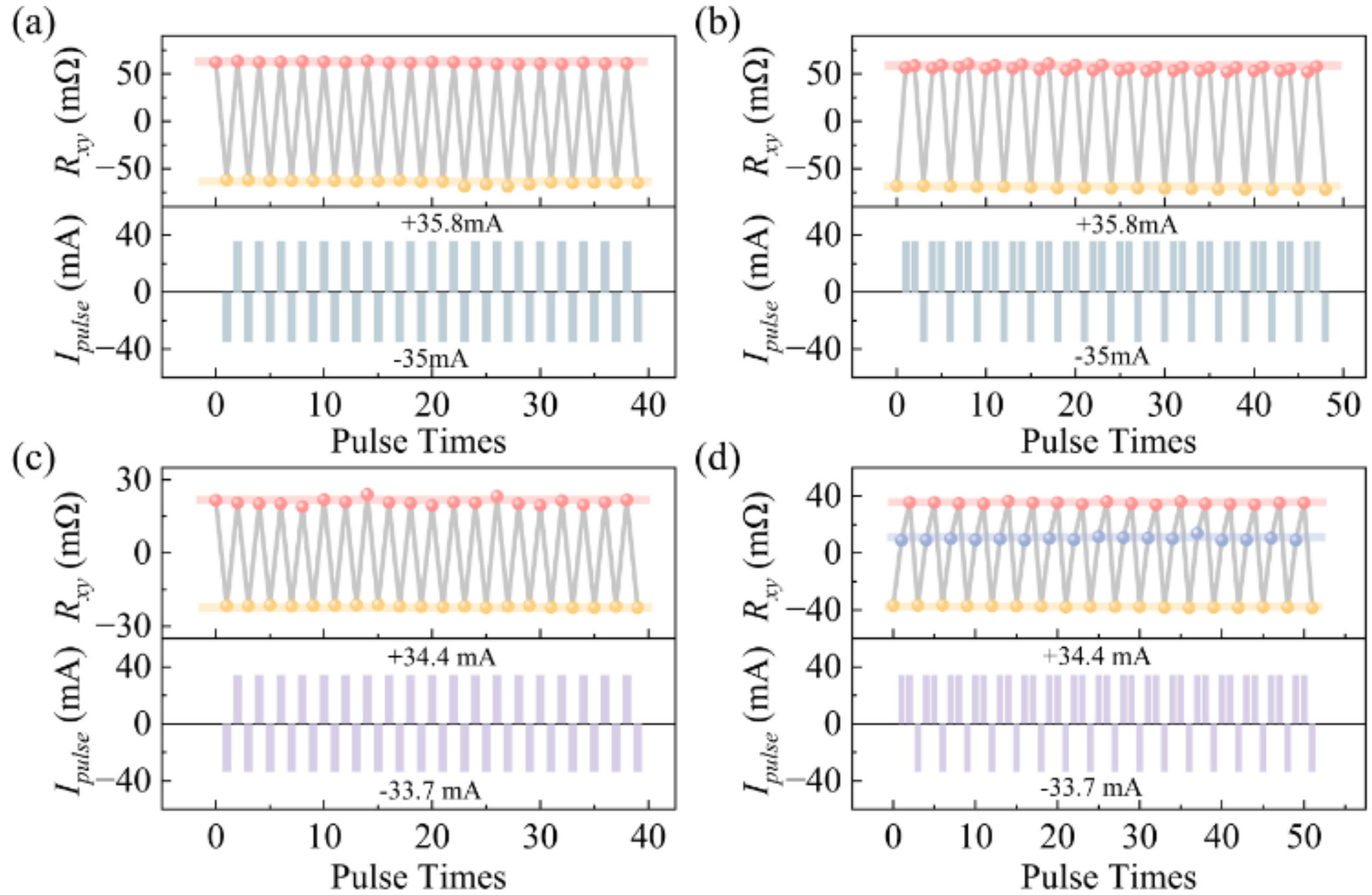

**Figure 4.** SOT switching modes under zero magnetic field. (a) Binary switching driven by a single write pulse of +35.8 mA and a reset pulse of -35 mA. (b) Binary switching driven by double consecutive write pulses of +35.8 mA and a reset pulse of -35 mA. (c) Partial switching driven by a single write pulse of +34.4 mA and a reset pulse of -33.7 mA. (d) Step-like switching exhibiting a distinct intermediate state driven by double consecutive write pulses of +34.4 mA and a reset pulse of -33.7 mA.

We use two pulse protocols to evaluate controlled current-driven operation. In the first, a single write current pulse $I_{write}$ is applied, followed by a reset current pulse $I_{reset}$ to initialize the device. In the second, two consecutive write pulses are applied before the reset operation. By tuning the magnitude of the $I_{write}$ pulses during these applications, the device can be driven to exhibit either direct switching or intermediate resistance states. To ensure that the magnetization stabilizes at a well-defined and reproducible resistance, the magnitude of $I_{reset}$ is matched to the corresponding $I_{write}$. The pulse conditions allow the device to be tuned between binary switching and multilevel switching at zero magnetic field. As shown in Figures 4(a) and 4(b), a large write current of +35.8 mA and a reset current of -35 mA produce a pronounced magnetization-reversal response with a relatively large switching amplitude. Both the single- and double-write-pulse schemes exhibit binary-like switching without stable intermediate states, indicating good cycling stability.

When the current is reduced to the critical range, specifically $I_{write}$ of +34.4 mA and an $I_{reset}$ of -33.7 mA, the switching behavior changes significantly. As illustrated in Figure 4(c), a single write pulse induces only partial magnetization switching. When two consecutive write pulses are applied, the switching becomes step-like, as shown in Figure 4(d). The first pulse drives a partial switching, while the second pulse drives the magnetization toward the final state. The gradient $Mn_{1.8-2.5}CoAl$ device can therefore be tuned between binary memory and analog synaptic behavior by adjusting the pulse amplitude and sequence.

### 3.4 Reconfigurable Logic Devices

Based on the field-free switching and accumulative characteristics, we demonstrate an all-electrically controlled reconfigurable spin logic scheme. Positive $R_{xy}$ corresponds to upward magnetization direction and is defined as logic output 1, while a negative $R_{xy}$ corresponds to a downward magnetization direction and is defined as logic output 0. By adjusting the magnitude and polarity of the input current pulses, four representative Boolean logic gates, including AND, NAND, OR, and NOR, are realized at zero external magnetic field. For each logic cycle, the device is first initialized by a reset pulse, followed by two consecutive input pulses representing the two operands. A low current of 5 mA is defined as logic input 0, while high current magnitudes are defined as logic input 1.

For the AND gate, shown in Figure 5(a), the operation relies on the accumulative switching property. A reset pulse of -33.7 mA initializes the device at output 0, while +34.4 mA represents logic input 1. A single logic 1 pulse drives the magnetization to an intermediate state where the resistance remains negative, corresponding to output 0. Only two consecutive logic 1 pulses enable the magnetization to switch across 0 and reach a positive resistance state, thereby producing an output of 1. As shown in Figure 5(d), for the OR gate, the device is initialized to output 0 by a reset pulse of -35 mA. The logic 1 input is set to +35.8 mA, which exceeds the deterministic switching threshold. As a result, any single logic 1 input is sufficient to switch the magnetization to the upward direction, yielding logic output 1. For the NOR gate in Figure 5(c), a reset pulse of +35 mA initializes the device to output 1. The logic 1 input is set to -35 mA, and logic 0 is -5 mA. In this case, any single logic 1 input reverses the magnetization downward to output 0. Similarly, for the NAND gate in Figure 5(b), the device is initialized to output 1 by a +34.5 mA reset pulse. With a logic 1 input of -34.3 mA, only two consecutive logic 1 pulses can pull the magnetization across zero to output 0.

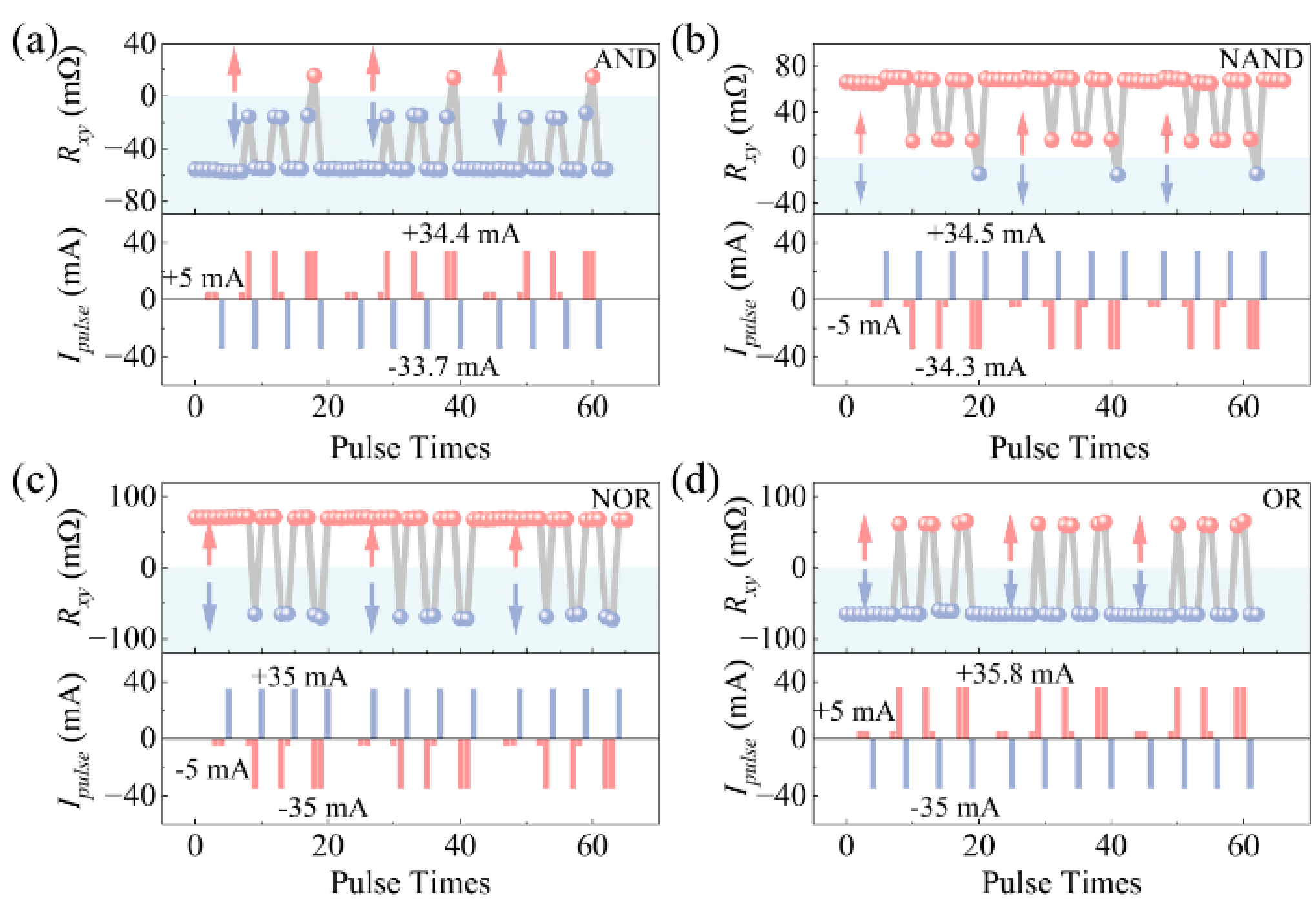

**Figure 5.** Experimental demonstration of reconfigurable Boolean logic gates within a single fixed physical unit under zero external magnetic field. (a) AND gate driven by a reset pulse of -33.7 mA and input 1 pulses of +34.4 mA. (b) NAND gate driven by a reset pulse of +34.5 mA and input 1 pulses of -34.3 mA. (c) NOR gate driven by a reset pulse of +35 mA and input 1 pulses of -35 mA. (d) OR gate driven by a reset pulse of -35 mA and input 1 pulses of +35.8 mA. Across all operations, the logic input 0 is represented by a 5 mA pulse with the corresponding polarity.

The reliability tests shown in Figure 5 confirm stable and repeatable switching between logic states. The magnetization switches stably between the logic 0 and logic 1 states without evident drift. In particular, the AND gate operations exhibit well-defined intermediate states, which enable clear threshold discrimination for logic evaluation. The intrinsic reconfigurability of SOT devices allows a single hardware unit to perform multiple logic functions by simply adjusting the input conditions.[31–35] This fully electrical and field-free operation eliminates the need for an externally applied magnetic field during logic operation and demonstrates a compact device-level implementation relevant to nonvolatile LIM architectures.

# 4 CONCLUSIONS

In this work, we investigate field-free SOT switching and reconfigurable spin logic in composition-engineered $Mn_xCoAl$/Pt Heusler heterostructures. Field-free current-induced magnetization switching is observed in the $Mn_{1.8\text{-}2.5}CoAl$ composition-gradient device but not in the uniform-composition controls. DMI-related measurements reveal a finite characteristic effective-field scale in the gradient device, indicating the involvement of chiral magnetization-reversal processes. The composition-gradient device also exhibits accumulative multistate switching with stable intermediate Hall-resistance states. By exploiting this multistate response and tailoring the amplitude, polarity, and sequence of current pulses, AND, OR, NAND, and NOR Boolean operations are realized within the same Hall-bar device. These results extend the functionality of the Mn-Co-Al Heusler system toward field-free magnetization control, multistate operation, and reconfigurable spin logic through vertical composition engineering.

# 5 ASSOCIATED CONTENT

**Supporting Information.** Extraction of the effective anisotropy field; estimation of the Pt-layer current density; evaluation of the DMI-related effective field; field-assisted switching and repeated-pulse measurements of the composition-uniform controls; symmetric Hall-bar geometry and photolithography pattern.

# 6 ACKNOWLEDGMENTS

The work was supported by the Research Start-up Funds of Hangzhou International Innovation Institute of Beihang University No. 2024KQ01 and CAS project for Young Scientists in Basic Research under Grant No. YSBR-030.

# 7 COMPETING FINANCIAL INTEREST

The authors declare that they have no conflict of interest.

# 8 DATA AVAILABILITY STATEMENT

The data that support the findings of this study are available from the corresponding author upon reasonable request.